\documentclass[aps,prl,twocolumn,floatfix]{revtex4}
\usepackage{graphicx}

\def\psii{\psi_{i}}
\def\psij{\psi_{j}}
\def\psiij{\psi_{ij}}
\def\psisq{\psi^2}

\def\psiibypsi{{\psii \over \psi}}
\def\psijbypsi{{\psij \over \psi}}
\def\psiijbypsi{{\psiij \over \psi}}

\def\EL{E_{\rm L}}
\def\ELi{E_{{\rm L},i}}
\def\ELj{E_{{\rm L},j}}

\def\Ebar{\bar{E}}
\def\Ei{\Ebar_{i}}
\def\Ej{\Ebar_{j}}
\def\Ebar{\bar{E}}

\def\NMC{N_{\rm MC}}

\def\MC{Monte Carlo}
\def\adiag{a_{\rm diag}}
\def\tcorr{T_{\rm corr}}
\def\intR{{\int {\rm d}^{3N}\!\!R\;}}

\def\beq{\begin{eqnarray}}
\def\eeq{\end{eqnarray}}

\begin{document}

\title{Energy and variance optimization of many body wave functions}

\author{C. J. Umrigar}
\affiliation{Theory Center and Laboratory of Atomic and Solid State Physics, Cornell University, Ithaca, NY 14853}
\author{Claudia Filippi}
\affiliation{Instituut Lorentz, Universiteit Leiden, Niels Bohrweg 2, Leiden, NL-2333 CA, The Netherlands}

\begin{abstract}
We present a simple, robust and efficient method for varying
the parameters in a many-body wave function to optimize the expectation
value of the energy.
The effectiveness of the method is demonstrated
by optimizing the parameters in flexible Jastrow factors, that include 3-body
electron-electron-nucleus correlation terms, for
the NO$_2$ and decapentaene (C$_{10}$H$_{12}$) molecules.
The basic idea is to add terms to the straightforward expression for the
Hessian of the energy that have zero expectation value,
but that cancel much of the statistical fluctuations for a finite
Monte Carlo sample.  The method is compared to what is currently the
most popular method for optimizing many-body wave functions, namely
minimization of the variance of the local energy.
The most efficient wave function is obtained by optimizing a linear combination
of the energy and the variance.
\end{abstract}

\date{\today}
\maketitle

Quantum Monte Carlo methods~\cite{FoulkesRMP01,NATObook,HammondLesterReynolds94} are
some of the most accurate and efficient methods for treating many body systems.
The success of these methods is in large part due to the flexibility in the form
of the trial wave functions that results from doing integrals by \MC.
Since the
capability to efficiently optimize the parameters in trial wave functions is crucial
to the success of both the variational \MC\ (VMC) and the diffusion \MC\ (DMC) methods,
a lot of effort has been put into inventing better optimization methods.

The variance minimization~\cite{varmin,UWW88} method has become the
most frequently used method for optimizing
many-body wave functions because it is far more efficient than {\it straightforward}
energy minimization.  The reason is that, for a sufficiently flexible variational
wave function, it is possible to lower the energy
on the finite set of Monte Carlo (MC) configurations on which the optimization is
performed, while in fact raising the true expectation value of the energy.
On the other hand, if the variance of the local energy is minimized, each
term in the sum over MC configurations is bounded from below by zero and the
problem
is far less severe~\cite{UWW88}.

Nevertheless, in recent years several clever methods have been invented that optimize
the energy rather than the
variance~\cite{Harju97,Snajdr99,Fahy99,FilippiFahy00,Rappe,Nightingale01,sr,SchautzFahy02,PrendergastFahy02,SchautzFilippi04}.
The motivations for this are four fold.  First, one
typically seeks
the lowest energy in either a variational or a diffusion
Monte Carlo calculation, rather than the lowest variance.
Second, although the variance minimization method has been used to optimize
both the Jastrow coefficients and the determinantal coefficients
(the coefficients in front of the determinants,
and in the expansion of the orbitals in a
basis, and the exponents in the Slater/Gaussian basis functions)
~\cite{UWW88,FilippiUmrigar96,HuangUmrigarNightingale97}, it takes many iterations to optimize the
latter and the optimization can get stuck in multiple local minima.
So, most authors have used variance minimization for the Jastrow parameters
only, where these problems are absent.
Third,
for a given form of the trial wave function, energy-minimized
wave functions on average yield more accurate values of other expectation values
than variance minimized wave functions do~\cite{Snajdr00}.
Fourth, the Hellman-Feynman theorem, combined with a variance reduction
technique~\cite{AssarafCaffarel}, can be used with energy-minimized wave functions
to compute forces on nuclei.

The various energy minimization methods are successful in varying degrees.
The generalized eigenvalue method of Nightingale and Melik-Alaverdian~\cite{Nightingale01}
is the most efficient choice for optimizing linear parameters, but for
nonlinear parameters they use variance minimization.
The effective fluctuation potential
method~\cite{Fahy99,FilippiFahy00,SchautzFahy02,PrendergastFahy02,SchautzFilippi04}
is the most successful
method for nonlinear parameters
and has been applied to optimizing the orbitals~\cite{FilippiFahy00,SchautzFilippi04} and
the linear coefficients in a multideterminantal wave function~\cite{SchautzFahy02,SchautzFilippi04},
and, has been extended to excited states~\cite{SchautzFilippi04}.
It is not straightforward to use this method to optimize Jastrow factors,
but Prendergast, Bevan and Fahy~\cite{PrendergastFahy02}
have formulated a version for periodic systems and have
optimized an impressively large number of parameters.
However, the method is complex and needs to be reformulated for finite systems.
The stochastic reconfiguration method~\cite{sr} is related to the effective
fluctuation potential method and is simpler but less efficient~\cite{SchautzFilippi04}.
The Newton method as implemented in Ref.~\onlinecite{Rappe} is the most straightforward
method but is inefficient and unstable.
The earlier methods~\cite{Harju97,Snajdr99} have been applied only to very
small systems or very few parameters.

The purpose of this letter is to show that it is possible to devise an
energy minimization method that
is simple, robust and efficient.
The method can be applied to optimizing many-body wave functions,
for both continuum and lattice problems.
The trick to doing this is to modify the straightforward expression for the
Hessian of the energy by adding a term that has
zero expectation value for an infinite MC sample, but that is nonzero and
cancels much of the statistical fluctuations for a finite MC sample.
Before we describe this in detail, we review the variance minimization method.

\vskip 2mm \noindent {\it Variance minimization:}\hskip 2mm
The parameters $c_i$ in a real-valued trial wave function $\psi$ are varied to
 minimize
the variance of the local energy,
\beq
\sigma^2
&=& {\intR \psisq (\EL-\Ebar)^2 \over \intR \psisq}
= \left\langle (\EL-\Ebar)^2 \right\rangle.
\eeq
where $\EL = {H \psi / \psi}$ is the local energy,
$\langle\cdot\rangle$ denotes a $\psi^2$-weighted expectation value,
and $\Ebar = \langle \EL \rangle$ is the expectation value of the energy.
The derivative of $\sigma^2$ with respect to the $i^{th}$ parameter, $c_i$, is given by
\beq
(\sigma^2)_i
&=& 2\Bigg[ \left\langle \ELi (\EL-\Ebar) \right\rangle \nonumber \\
&&+ \left\langle \left({\psii \over \psi} - \left\langle{\psii \over \psi}\right\rangle \right) (\EL-\Ebar)^2 \right\rangle
\Bigg]
\nonumber \\
&=& 2\Bigg[ \left\langle \ELi (\EL-\Ebar) \right\rangle
+ \left\langle {\psii \over \psi} \EL^2 \right\rangle
- \left\langle \psiibypsi \right\rangle
\left\langle \EL^2 \right\rangle \nonumber \\
&&-2 \Ebar \left\langle \psiibypsi (\EL-\Ebar) \right\rangle
\Bigg],
\label{sigma_der}
\eeq
where subscript $i$ denotes derivative with respect to $c_i$.

Since the variance minimization method can be viewed as a fit of the
local energy on a fixed set of Monte Carlo configurations~\cite{UWW88}, an alternative
expression follows from ignoring the change of the wave function:
\beq
\!\!\!\!\!\!(\sigma^2)_i \!&=&\! 2 \left\langle \ELi (\EL-\Ebar) \right\rangle
\!=\! 2 \left\langle (\ELi-\Ei) (\EL-\Ebar) \right\rangle
\eeq

Then the usual Levenberg-Marquardt approximation~\cite{recipes} to the Hessian
matrix is given by
\beq
\!\!\!\!\!\!
(\sigma^2)_{ij} &\!\!=&\!\! 2 \left\langle (\ELi-\Ei) (\ELj-\Ej) \right\rangle
\nonumber \\
&\!\!=&\!\! 2\left( \left\langle \ELi\ELj \right\rangle
- \Ei \left\langle \ELj \right\rangle
- \Ej \left\langle \ELi \right\rangle
+ \Ei\Ej \right)
\label{LevenbergMarquardt}
\eeq
This Hessian is positive definite by construction.

\vskip 2mm \noindent {\it Energy minimization:} \hskip 2mm
The elements of the gradient are
\beq
\Ei
&\!\!=&\!\! \left\langle \psiibypsi \EL + { H \psii \over \psi}
-2 \Ebar \psiibypsi \right\rangle
\label{first_deriv_nonherm}
\\
&\!\!=&\!\! 2\left\langle \psiibypsi (\EL - \Ebar) \right\rangle
\;\;\;\; {\rm (by\;Hermiticity)}.
\label{first_deriv}
\eeq
We note that the step from
Eq.~\ref{first_deriv_nonherm} to Eq.~\ref{first_deriv} was made
not just in the interest of simplicity, but more importantly
because the expression in Eq.~\ref{first_deriv} has zero fluctuations
in the limit that $\psi$ is an exact eigenstate, whereas the expression
in Eq.~\ref{first_deriv_nonherm} has large fluctuations.

Taking the derivative of Eq.~\ref{first_deriv}, the Hessian is
\beq
\Ebar_{ij}
&=& 2 \Bigg[
\left\langle \left( \psiijbypsi + {\psii\psij\over \psisq} \right) (\EL-\Ebar) \right\rangle \nonumber \\
&&
-\left\langle \psiibypsi \right\rangle \Ej
-\left\langle \psijbypsi \right\rangle \Ei
+ \left\langle \psiibypsi \ELj \right\rangle \Bigg].
\label{rappe}
\eeq
This is nothing more than a rearrangement of terms in the Hessian in Ref.~\cite{Rappe}.
We now make two changes to the above expression.
First, we note that the last term is not symmetric in $i$ and $j$
when approximated by a finite sample, whereas
the true Hessian of course is symmetric.  So, we symmetrize it.
This change does not significantly alter the efficiency of the method,
but it does have the advantage that the eigensystem is real.
Next, we note that Eq.~\ref{first_deriv} and all except the last term in
Eq.~\ref{rappe} are in the form of a covariance,
($\langle ab \rangle - \langle a \rangle \langle b \rangle$).
The fluctuations of $\langle ab \rangle - \langle a \rangle \langle b \rangle$
are in most cases smaller than those of $\langle ab \rangle$, (e.g. if
$a$ and $b$ are weakly correlated), and, they are much smaller if
$\sqrt{\langle a^2 \rangle - \langle a \rangle ^2} \ll \vert \langle a \rangle \vert$
and $a$ is not strongly correlated with $1/b$.
Since the Hamiltonian is Hermitian it follows, as also noted
in Ref.~\cite{Rappe}, that
$ \langle \ELj \rangle = 0$.
Hence, an alternative symmetric expression~\cite{Sorella} for the Hessian, written entirely in terms
of covariances, is:
\beq
\Ebar_{ij} &=&
2 \Bigg[
\left\langle \left( \psiijbypsi + {\psii\psij\over \psisq} \right) (\EL-\Ebar) \right\rangle
-\left\langle \psiibypsi \right\rangle \Ej \nonumber \\
&&
-\left\langle \psijbypsi \right\rangle \Ei \Bigg] 
+ \left\langle \psiibypsi \ELj \right\rangle
+ \left\langle \psijbypsi \ELi \right\rangle \nonumber \\
&&
- \left\langle \psiibypsi \right\rangle \left\langle \ELj \right\rangle
- \left\langle \psijbypsi \right\rangle \left\langle \ELi \right\rangle.
\label{Hessian}
\eeq
The additional terms we have added in have zero expectation value
for an infinite sample but, in practice, cancel most of the fluctuations in the existing
terms for a finite sample, making the method vastly more efficient.
Note also that $\Ebar_{ij}$ in Eq.~\ref{rappe}, evaluated on a finite sample, is not
invariant under renormalization of the wave function by a parameter-dependent constant
but $\Ebar_{ij}$ in Eq.~\ref{Hessian} is.

We note that Eqs.~\ref{first_deriv} and \ref{Hessian} are
{\it not} the gradient and the Hessian of the energy estimated on
the particular finite set of sampled points.
In fact, any method that attempts to
minimize the energy, by minimizing the energy evaluated on a finite sample
of \MC\ points, is bound to require a very large sample and therefore be
highly inefficient for
the reason discussed in the introduction.
Our modifications of the straightforward expressions for the gradient
and Hessian are similar in spirit to the work of
Nightingale and Melik-Alaverdian~\cite{Nightingale01}.
A straightforward minimization of the energy on a \MC\ sample
results in a symmetric Hamiltonian matrix, but they derive a
nonsymmetric Hamiltonian matrix that yields exact parameters
from a finite sample in the limit that the basis functions span an invariant subspace.

\vskip 2mm \noindent {\it Newton method:} \hskip 2mm
In both the energy and the variance minimization methods, the
gradient, {\bf b}, and the Hessian, {\bf A}, are used to update the
variational parameters, ${\bf c}$, using Newton's method,
${\bf c}_{\rm next} = {\bf c}_{\rm current} - {\bf A}^{-1}{\bf b}$.

Note that if we are far away from the minimum, or if
the number of Monte Carlo samples,
$\NMC$, is small, then the Hessian of Eq.~\ref{Hessian} need not be positive
definite, whereas
the approximate Hessian of Eq.~\ref{LevenbergMarquardt} is always
positive definite.
Further, even for positive definite Hessians, the new parameter values
may make the wave function worse if
one is not sufficiently close to the minimum for the quadratic approximation
to hold or if the approximate Hessian of Eq.~\ref{LevenbergMarquardt} is not
sufficiently accurate.
Hence, we determine the eigenvalues of the Hessian and add to the diagonal
of the Hessian the negative of the most negative eigenvalue (if one
exists) plus a constant $a_{\rm diag}$.
This shifts the eigenvalues by the added constant.
As $a_{\rm diag}$ is increased,
the proposed parameter changes become smaller and rotate from the
Newtonian direction to the steepest descent direction.
As an aside, we note that for the form of the wave functions used
and the molecules studied here,
we find that the eigenvalues of the Hessians of Eqs.~\ref{Hessian}
and \ref{LevenbergMarquardt} span 11 orders of magnitude when the
parameters are close to optimal.

\vskip 2mm \noindent {\it Results:} \hskip 2mm
We have tested the methods on NO$_2$ and
the excited $^1B_u$ state of decapentaene (C$_{10}$H$_{12}$)
using a nonlocal pseudopotential to remove core electrons.
We optimize the parameters in a flexible Jastrow factor~\cite{FilippiUmrigar96} that contains
electron-electron, electron-nucleus and electron-electron-nucleus terms,
making a total of 43 free parameters.
The starting Jastrow is a crude electron-electron
Jastrow of the form $\exp({br \over 1+r})$, where $b$ is set by the
cusp conditions for antiparallel- and parallel-spin electrons.

\begin{figure}[htb]
\includegraphics[width=3.5in,clip]{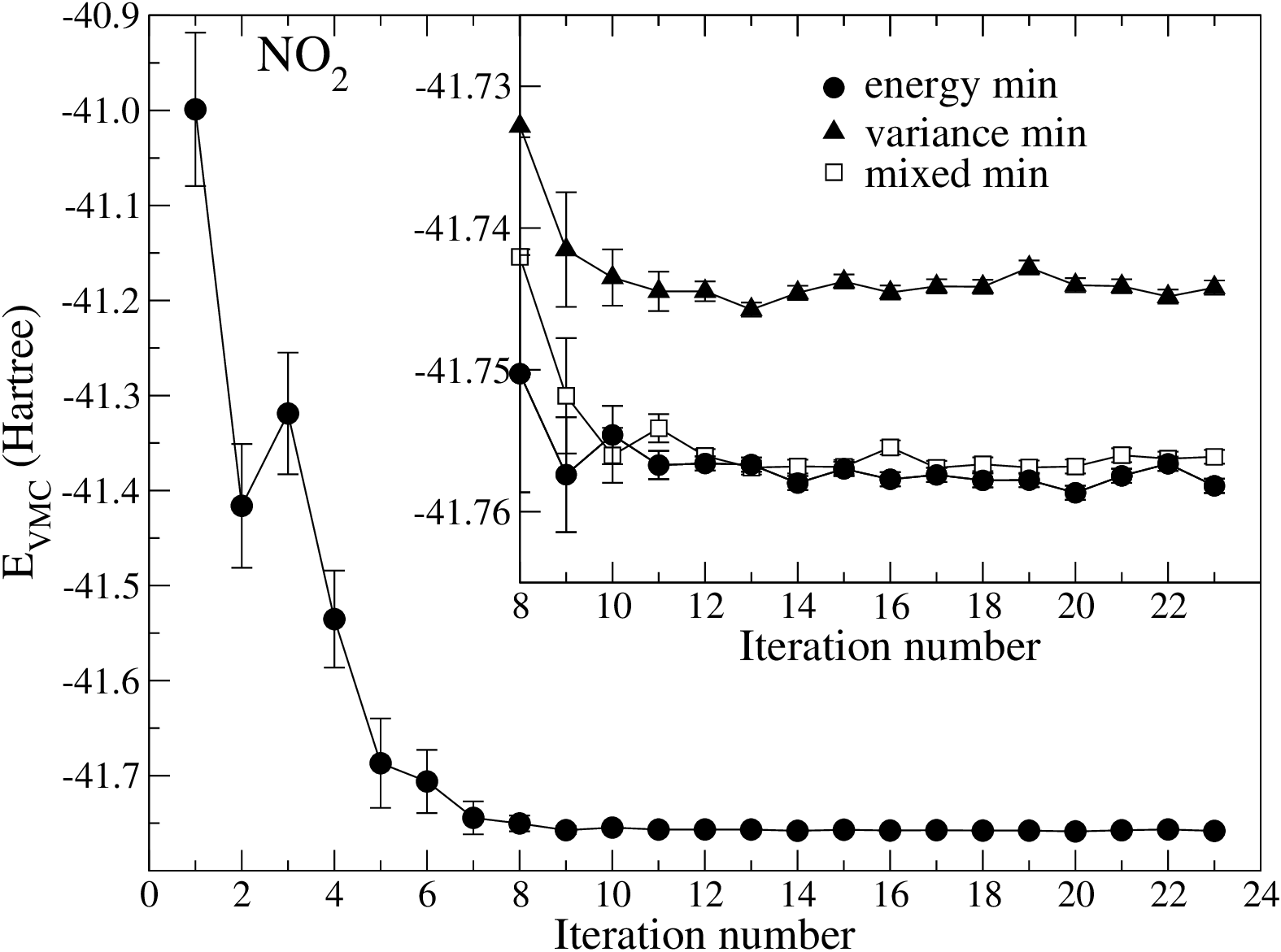}
\caption{Energy of NO$_2$ versus iteration number for energy
minimization.
Inset: the later iterations on an expanded scale and also
the energies from minimizing the variance and minimizing the linear combination.
The linear combination yields almost as good
an energy as energy minimization.
}
\label{energy_no2}
\end{figure}

\begin{figure}[htb]
\includegraphics[width=3.5in,clip]{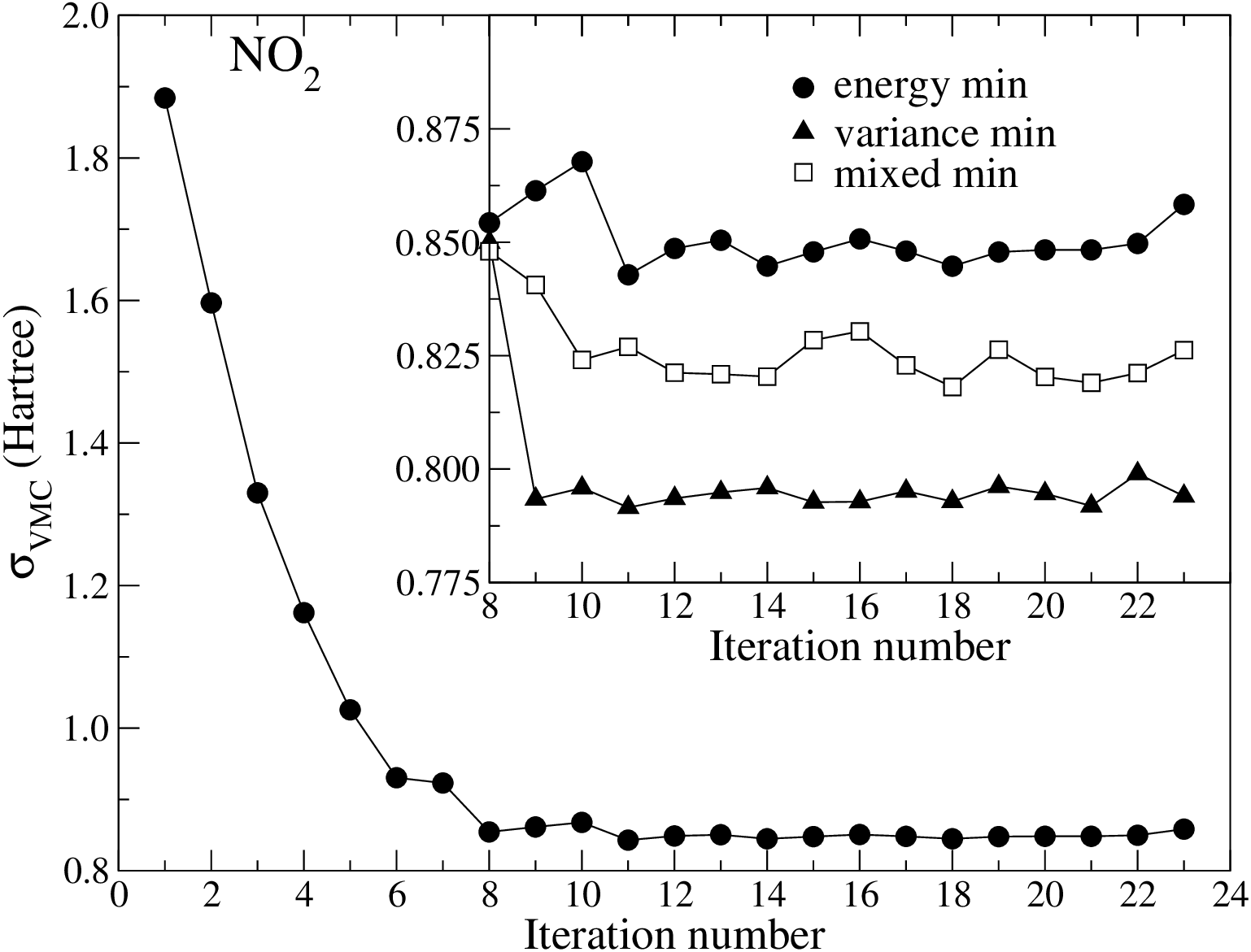}
\caption{Same as Fig.~\ref{energy_no2} but for the rms fluctuations of the local energy,
$\sigma$, rather than the energy.
The linear combination $\sigma$ is half way between
those from energy minimization and variance minimization.}
\label{sigma_no2}
\end{figure}

\begin{figure}[htb]
\includegraphics[width=3.5in,clip]{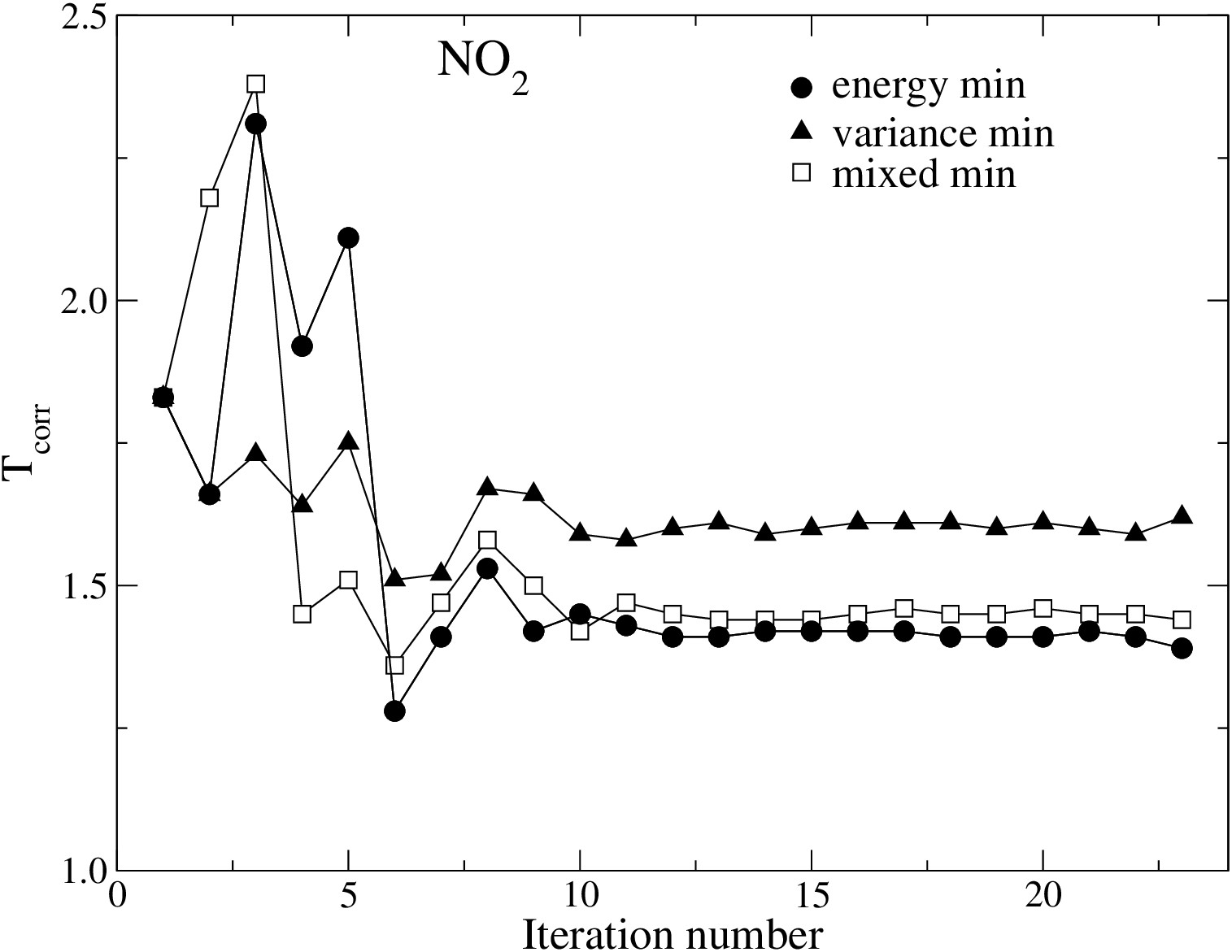}
\caption{The autocorrelation time, $T_{corr}$,
of NO$_2$ versus iteration number.
Energy minimization gives the smallest $T_{\rm corr}$, variance minimization the largest,
and, the mixed minimization a value that is close to that from energy minimization.
}
\label{tcorr_no2}
\end{figure}

\begin{figure}[htb]
\includegraphics[width=3.5in,clip]{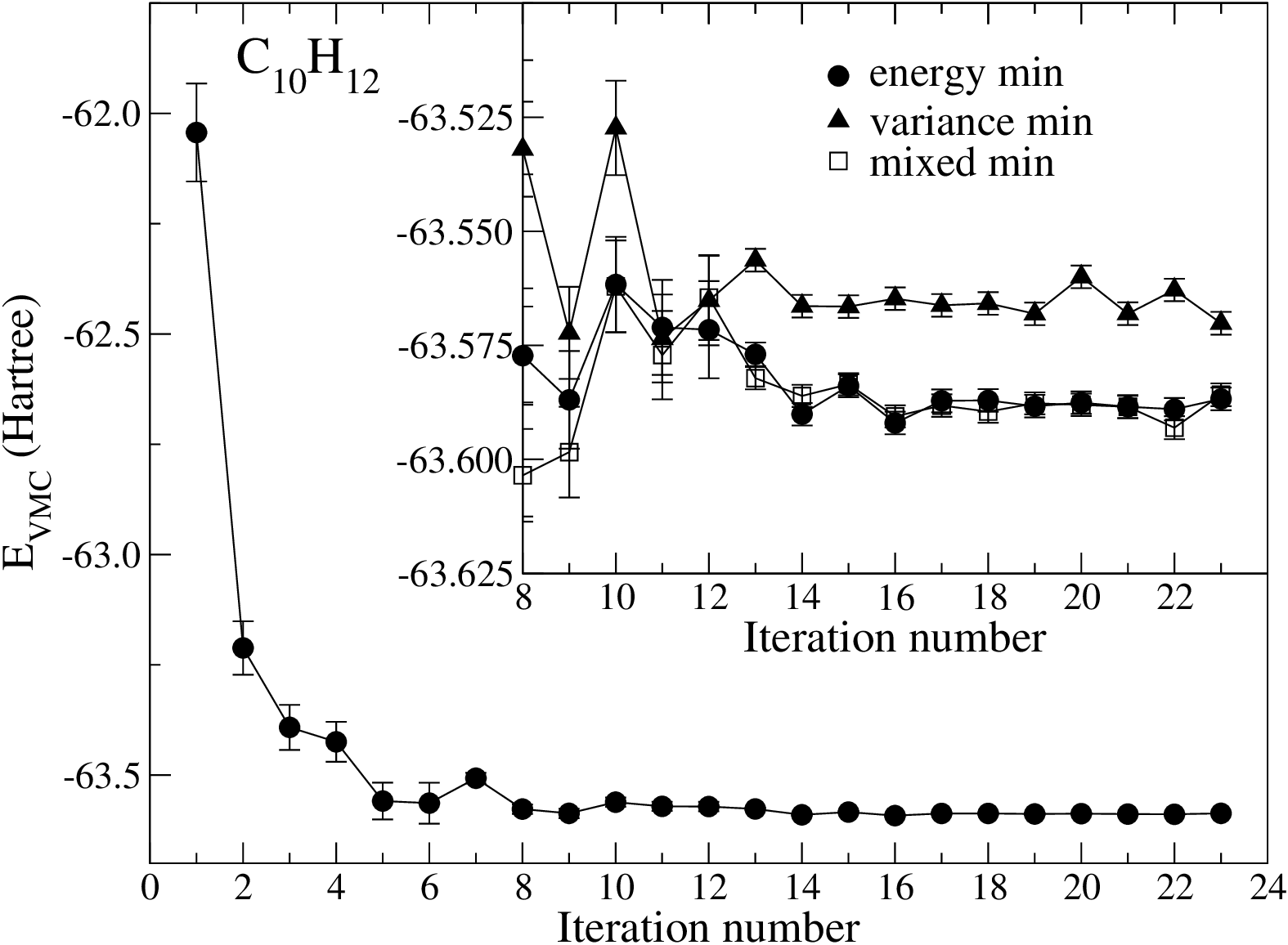}
\caption{Same as Fig.~\ref{energy_no2} but for decapentaene (C$_{10}$H$_{12}$).}
\label{energy_decapentaene}
\end{figure}

In Fig.~\ref{energy_no2}, we plot the energy, and, in Fig.~\ref{sigma_no2}
the root mean square fluctuations of the local energy, $\sigma$,
of NO$_2$ as a function of the
iteration number as we energy optimize the 43 free parameters in the Jastrow.
The first 6 iterations employ a very small MC sample,
$\NMC=1000$, and $a_{\rm diag}=0.2$.  For each of the next 6 iterations
we increase $\NMC$ by a multiplicative factor of 4 and decrease
$a_{\rm diag}$ by a multiplicative factor of 0.1.  The remaining 11 iterations
are performed with the values at the end of this process, namely,
$\NMC=4,096,000$, and $a_{\rm diag}=2\times10^{-7}$.
(Setting $\adiag=0$ would work equally well for these iterations.)
The first few iterations are extremely fast due to the small value
of $\NMC$ and achieve most of the optimization.
In the insets we show the later iterations on an expanded scale, and
also the energies and $\sigma$ from minimizing the variance
(using Eqs.~\ref{sigma_der} and \ref{LevenbergMarquardt}) and from minimizing
a linear combination, with the variance having a weight
of 0.05 and the energy a weight of 0.95.
Of course, the variance-minimized wave functions have a lower $\sigma$
and the energy-minimized wave functions a lower energy.
The mixed-minimization wave functions
have an energy that is almost as good as that of the energy-minimized
wave functions, and, a $\sigma$ that is in between.

The computational time required to reduce the statistical error to
a given value is proportional to $\sigma^2\tcorr$, where $\tcorr$ is
the autocorrelation time of the energy as defined in Ref.~\cite{accel_metrop}.
One can argue that in DMC the energy
minimized wave functions will have a smaller $\tcorr$
than variance minimized wave functions,
since both $\sigma$ and $\tcorr$ serve to lower the DMC
energy relative to the variational energy.
In Fig.~\ref{tcorr_no2}, we show $\tcorr$ for each of the three
methods.  We see that the energy minimized wave function has a smaller
value of $\tcorr$ than the variance minimized wave function, even in VMC.
The mixed minimization wave function has a $\tcorr$
that is close to that of the energy minimized wave function.
The value of $\sigma^2\tcorr$ for the variance, energy and mixed
optimizations is 1.08, 1.03 and 0.98 H in VMC, and,
3.21, 2.87 and 2.75 H in DMC using a time-step of 0.05 H$^{-1}$,
where the last digit in $\sigma^2\tcorr$ is uncertain.
Hence, the wave functions obtained from the mixed optimization are the
most efficient ones.

We note that $E$ and $\sigma$ are fully converged in 12 iterations.
In fact, it is possible to converge them in 4-5 iterations
if we use from the outset a larger value for $\NMC$ and reduce the
value for $a_{\rm diag}$ more rapidly.
However, it is more computationally efficient to start the optimization
by performing several iterations with a small $\NMC$.

In Fig.~\ref{energy_decapentaene} we plot the energy of the excited $^1B_u$ state
of a larger molecule, decapentaene (C$_{10}$H$_{12}$), as a function of iteration number.
For the first 6 iterations we optimize just the 13 parameters
in the electron-nucleus and the electron-electron Jastrows, and, optimize
the full set of 43 parameters starting from iteration 7.  As in the case
of NO$_2$, we employ $\NMC=1000$ and $\adiag=0.2$
during the first six iterations.  The next six are performed with
$\NMC=16000$ and the final 11 iterations are performed with
$\NMC=256000$ and $\adiag=2\times 10^{-5}$.
The results are similar to those for NO$_2$, and so in the interest
of brevity we omit plots for $\sigma$ and $\tcorr$.

It is remarkable that most of the optimization can be done with
as few as 1000 MC configurations.  In contrast, if Eq.~\ref{rappe}
is used for the Hessian, then the fluctuations are much larger
and the method becomes unstable for the molecules treated here
even if we increase the number of \MC\ configurations, $\NMC$, by a factor
of a thousand to $10^6$ configurations.  (We can make it stable
by increasing substantially also the value of $a_{\rm diag}$, but this
increases the number of iterations needed to converge.)
Hence, the simple change going from Eq.~\ref{rappe} to Eq.~\ref{Hessian},
that entails no additional computational cost,
results in a gain in efficiency of at least three orders of magnitude.

\vskip 2mm \noindent {\it Acknowlegements:}
We thank Peter Nightingale for valuable discussions.
Supported by NSF (DMR-0205328),
NASA, Sandia National Laboratory
and Stichting voor Fundamenteel Onderzoek der Materie (FOM).


\begin{thebibliography}{99}

\bibitem{FoulkesRMP01}
W.~M.~C.~Foulkes, L.~Mitas, R.~J.~Needs, and G.~Rajagopal, Rev.~Mod.~Phys.\
{\bf 73}, 33 (2001).

\bibitem{NATObook}
{\it Quantum Monte Carlo Methods in Physics and
Chemistry}, edited by M. P. Nightingale and C. J. Umrigar,
[NATO ASI Ser. C. {\bf 525} 101, 1999].

\bibitem{HammondLesterReynolds94}
B.L. Hammond, W.A. Lester and P.J. Reynolds,
{\it Monte Carlo Methods in Ab Initio Quantum Chemistry}, (World Scientific 1994).

\bibitem{varmin}
The idea of minimizing the variance of the local energy goes back to at least
1935,  J.H. Bartlett, J.J. Gibbons and C.G. Dunn, Phys. Rev. {\bf 47}, 679 (1935).
It was first used in quantum Monte Carlo by
R.L. Coldwell, Int. J. Quant. Chem. Symp., {\bf 11}, 215 (1977).

\bibitem{UWW88} C. J. Umrigar, K. G. Wilson, J. W. Wilkins, Phys. Rev. Lett. {\bf 60}, 1719 (1988);
C.J. Umrigar, K.G. Wilson and J.W. Wilkins, in
{\it Computer Simulation Studies in Condensed Matter Physics: Recent Developments},
ed. by D.P. Landau K.K. Mon and H.B. Sch\"uttler,
Springer Proc. Phys. (Springer, Berlin 1988);
C.J. Umrigar, {\it Int. J. Quant. Chem.} {\bf 23}, 217 (1989).

\bibitem{Harju97} A. Harju, B. Barbiellini, S. Siljam\"aki, R. M. Nieminen, and G. Ortiz,
Phys. Rev. Lett. {\bf 79}, 1173 (1997).

\bibitem{Snajdr99}
Martin Snajdr, Jason R. Dwyer, and Stuart M. Rothstein,
J. Chem. Phys. {\bf 111}, 9971 (1999), erratum {\bf 114}, 6960 (2001).



\bibitem{Rappe} Xi Lin, Hongkai Zhang and Andrew M. Rappe,
J. Chem. Phys., {\bf 112}, 2650 (2000);
Myung Won Lee, Massimo Mella, Andrew M. Rappe, arXiv:physics/0411209.

\bibitem{Nightingale01} M. P. Nightingale and Melik-Alaverdian,
Phys. Rev. Lett., {\bf 87}, 043401 (2001).


\bibitem{Fahy99} S. Fahy, in Ref.~\cite{NATObook}.
\bibitem{FilippiFahy00} Claudia Filippi and Stephen Fahy, J. Chem. Phys., {\bf 112}, 3523 (2000).
\bibitem{SchautzFahy02} Friedemann Schautz and Stephen Fahy, J. Chem. Phys., {\bf 116}, 3533 (2002).
\bibitem{PrendergastFahy02} David Prendergast, David Bevan and Stephen Fahy, Phys. Rev. B, {\bf 66}, 155104 (2002);
\bibitem{SchautzFilippi04} Friedemann Schautz and Claudia Filippi, J. Chem. Phys., {\bf 120}, 10931 (2004).

\bibitem{sr} Sandro Sorella, Phys. Rev. B, {\bf 64}, 024512 (2001);
Michele Casula and Sandro Sorella, J. Chem. Phys., {\bf 119}, 6500 (2003).

\bibitem{FilippiUmrigar96}
Claudia Filippi and C.J. Umrigar, J. Chem. Phys., {\bf 105}, 213 (1996).
Our Jastrow is related to the one here.

\bibitem{HuangUmrigarNightingale97}
Chien-Jung Huang, C.J. Umrigar and M.P. Nightingale, J. Chem. Phys. {\bf 107}, 3007 (1997).

\bibitem{Snajdr00}
Martin Snajdr and Stuart M. Rothstein, J. Chem. Phys. {\bf 112}, 4935 (2000).

\bibitem{AssarafCaffarel}
R. Assaraf and M. Caffarel, J. Chem. Phys. {\bf 113}, 4028 (2000);
{\it ibid} {\bf 119}, 10536 (2003).

\bibitem{recipes} See e.g. W.H. Press, B.P. Flannery, S.A. Teukolsky, and W.T. Vetterling,
{\it Numerical Recipes}, (Cambridge University Press, Cambridge 1992).

\bibitem{Sorella} After submission we learned that Eq.~\ref{Hessian} has been
independently derived by S. Sorella, to be submitted to cond.-mat.

\bibitem{accel_metrop} C.J. Umrigar, Phys. Rev. Lett. {\bf 71}, 408 (1993);
C. J. Umrigar, in Ref.~\cite{NATObook}.

\end{thebibliography}
\end{document}